\documentclass[review]{elsarticle}

\usepackage{hyperref}
\usepackage{booktabs,bm}
\usepackage{footnote}
\usepackage[export]{adjustbox}
\usepackage{amsmath}

\journal{Infectious Disease Modelling}

\begin{document}

\begin{frontmatter}

\title{The fallacy of herd immunity}

\author[UC]{Carlos Hernandez-Suarez\corref{cor1}}

\address[UC]{Facultad de Ciencias, Universidad de Colima, Bernal Diaz del Castillo 340, Colima, Colima, 28040, MEXICO}

 \cortext[cor1]{Corresponding author}

\begin{abstract}
Here we show that herd immunity is not a measure we can implement, trust, control or depend on, and that waiting for herd immunity to play a role is tantamount to hope for the best. Herd immunity is not some protection, as it has been referred to, but a reduction in the chance of infection that a population can achieve only if more infections occur, which is paradoxical. The discussion on when there will be enough herd immunity for the COVID-19 epidemics to halt, delivers the public and policy makers the wrong message that we can wait for something to happen to relax containment measures. Here we review a non-orthodox construction of an epidemic that unveils what \textit{herd immunity} really means, that is easy to understand by non-experts.

\end{abstract}

\begin{keyword}
Herd immunity \sep Epidemic size \sep Basic reproductive number \sep Urn models \sep $R_0$ 
\end{keyword}

\end{frontmatter}


\section{Introduction}
The term \textit{herd immunity} has been recently referred to as some policy that can be implemented, when in fact, is the opposite: it represents total inaction. Countries like Sweden have opted for this \cite{conyon2020lockdowns} and in the beginning the United Kingdom planned to used it as the official strategy \cite{Horton:2020ab}. In here we discuss what can be expected from this inaction. Herd immunity is not some kind of immunity at all. A fraction $f$ of the population will escape infection by pure chance and herd immunity does not change that fraction. All the discussion on when there will be enough \textit{herd immunity} \cite{kwok2020herd} for the epidemics to halt makes people believe that we can wait for something to happen that will reduce our risk of infection. In fact, the only way to achieve \textit{heard immunity} is if more people is infected. We should all be aware that \textit{herd immunity} is not the same as \textit{biological immunity} \cite{10.1093/cid/cir007}.

To understand what herd immunity really means it is useful to rely on a simple epidemic model that is very intuitive and can easily be understood by non-experts and policy makers. The goal of this paper is to exhibit that the words ``immunity'' and ``protection'' are not part of what we should understand for \textit{herd immunity}.

\section{Alternative construction of an epidemic model}
The simplest and more intuitive epidemic model is the SIR family of models model \cite{kermack1927contribution} that basically divide the population in several stages: susceptible, infectious or removed. Those removed are individuals that recovered or died from the disease, playing no role in the transmission. A variation is the SEIR model, that adds a `latent state'. In here, we are interested in the epidemic size, that is, the total number of infected at the end of the epidemic and we will assume random mixing, which implies that an infected individual may infect any susceptible. The goal of this paper is not to propose a model nor to forecast the epidemic size of the current pandemics, but to explain what is \textit{herd immunity} and for this, the SIR model will do. 

The most important parameter in the dynamics of an epidemic is $R_0$, the \textit{basic reproductive number}, defined it as ``The average number of infections caused by an infected individual when it is introduced to a population of susceptibles'' \cite{MR1057044}. We will construct an epidemic in such a way that the definition of $R_0$ is more evident and intuitive. 

\begin{figure}[htp]
\begin{center}
\includegraphics[width=1.1 \textwidth,left]{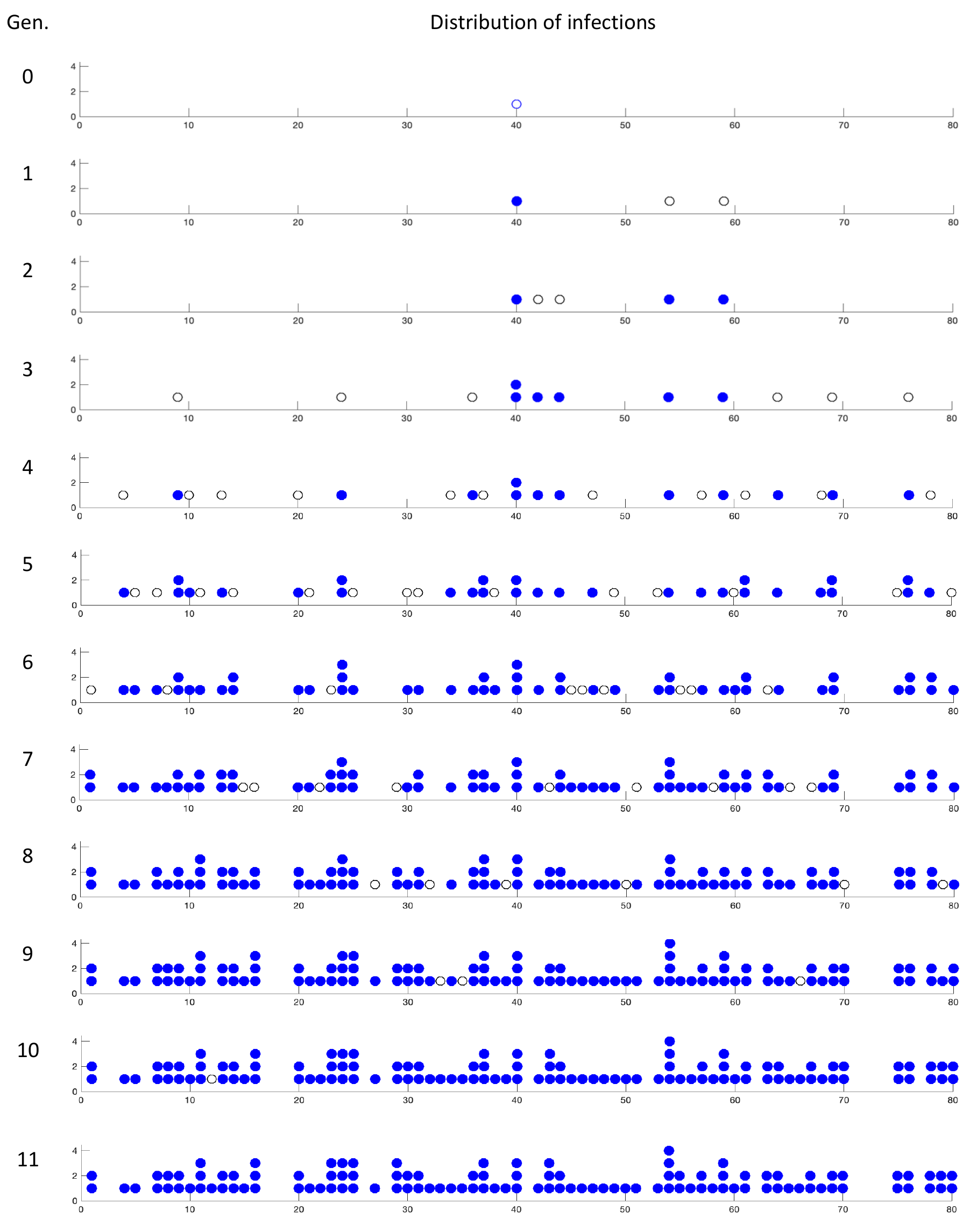}
\caption{The progression of an epidemic with $R_0=2$ in a population with $N=80$ individuals. The epidemics halted at generation 11, when the last generation of balls fell in already occupied urns.}
\label{F1}
\end{center}
\end{figure}

For simplicity, assume a closed population of size $N$ and an SIR epidemic running in the population with $R_0=2$. The construction we will use is based on placing balls in urns, that is, urn models, developed in \cite{hernandez2009applications}, in which we basically place balls at random in several urns, and each of the urns has unlimited capacity. The only result we need from urn theory is this: if we place $b$ balls at random in $N$ urns, the average or expected number of urns that will receive at least one ball (occupied urns) when $N$ and $b$ are large, is approximately:

\begin{equation*}
E[X] = N(1- e^{-b/N})
\label{eq1}
\end{equation*}
We construct the epidemic as follows, and we will refer to Figure (\ref{F1}) to support the explanation. There are three kinds of urns: empty, that are the susceptible urns, occupied with a single white ball, that correspond to the infectious urns, and occupied with one or more blue balls, that correspond to the recovered urns. In our example $N=80$. 

At generation 0, we place one white ball at random in one of the $N$ urns. This is our first infected urn, that is infectious. In this case, the infectious urn is urn 40.

At the next generation, this first occupied urn throws to the air a random number of \textit{white} balls from some discrete distribution with positive support and mean $R_0$ (any discrete distribution with these characteristics can be used, the result is the same \cite{hernandez2009applications}). All balls falling in empty urns become white, while those that fall in occupied urns becomes blue.

Every new generation is constructed using only the number of urns containing a single white ball in the previous generation, as explained before. Once an urn containing a single white ball throws to the air all its balls, the urn becomes recovered, that is, all of the balls that it contained become blue.

In generation 1, we can see that the first individuals threw away two balls and occupied urns 54 and 59. In generation 2, these urns recovered but before they threw to the air two balls that landed in urns 42 and 44, and so on.

 This construction allows to see what is the true meaning of $R_0$: is the average number of balls thrown by each infected urn. Thus, it is the average \textit{number of attempts}, that is, a measure of infection pressure. Clearly, all balls thrown to the air by the first infected will very likely land in an empty urn, that is why it is easy to define $R_0$ as ``The average number of infections caused by an infected individual when it is introduced to a population of susceptibles'' \cite{MR1057044}.

We can see that at generation 10, there was only one individual infectious (individual 12) and when this individual threw to the air the corresponding balls, all of them landed in occupied urns, and there were no more new infections and the epidemic halted (this individual threw two balls that landed in urns 29 and 55). 

Now, what halted the epidemic was \textit{herd immunity}.

The speed of growth of the epidemic is the speed at which new infections are acquired. At the beginning is low because although there are plenty of susceptibles, there are few infectious. At the end is again slow, because there are plenty of infectious, but few susceptibles. The peak is somewhere in between, when the product of the number of infectious and the number of susceptibles is maximum. In our model, this is achieved when the number of infectious equals the number of susceptibles. After this point, the risk of infection of \textit{those that have not been infected} reduces. In Figure (\ref{F1}) this is achieved in generation 4, when there are 11 infectious (white balls) and 58 susceptibles (empty urns). After generation 4, \textit{herd immunity} starts to play a role, reducing the number of empty urns that can be `infected'.

\section{The epidemic size}

First, observe that it is unlikely, under the underlying assumptions and $R_0=2$, that all the urns will be occupied. The explanation has the same background than the classical \textit{birthday problem} in which we only need 23 individuals to reach a probability of 50\% that at least two of them have the same birthday. If we place 240 balls at random in the 80 urns, that is, three times the number of urns, the probability that all of the urns will be occupied is less than $1.5\%$. Thus, some may escape infection because of pure luck, and not because of some policy or strategy or some kind of collective immunity or protection.

Thus, a relevant question is: what is the average number of infections that will occur? Hernandez \citep{hernandez2009applications} already showed that, if $I$ is the total number of individuals that will be infected at the end of the epidemic, that is, the epidemic size, then, from equation (\ref{eq1}), the expected final number of infections is:

\begin{equation*}
E[I] = E[N(1-e^{-b/N})]
\end{equation*}
which can be approximated using the first term of the Taylor series expansion with:
\begin{equation*}
E[I] = N(1-e^{-E[b]/N})
\end{equation*}
but $E[b]$ is the expected total number of balls thrown, which is clearly $E[I]R_0$, thus:
\begin{equation}
E[I] = N(1-e^{-E[I] R_0/N})
\label{ex}
\end{equation}
which has no closed solution but can easily be approximated numerically. In our example with $N=80$ and $R_0=2$, the solution of $E[I]$ in \ref{ex}, that is,  the expected number of infected is $63.6$, close to the observed value $64$ in the simulation of Figure (\ref{F1}).

\section{Conclusions}
\textit{Herd immunity} has been defined as ``the indirect protection from infection conferred to susceptible individuals when a sufficiently large proportion of immune individuals exist in a population'' \citep{randolph2020herd}, but in here, we have clarified that there is no such protection, but it is us waiting for the enemy to run out of bullets while shooting at us. That is, there is no herd immunity we can rely on. There is herd immunity that will happen, naturally. It is not a complex concept as it has been argued but the result of chance \cite{Rashid:2012aa}. The suggested implication of the term as that it is ``the risk of infection among susceptible individuals in a population is reduced by the presence and proximity of immune individuals'' \cite{10.1093/cid/cir007}  is erroneous. 

Vaccines do not offer \textit{herd immunity}. They reduce the fraction of susceptible so that $R_0 < 1$. The word immunity is reserved for some kind of protection to an agent when an individual is subjected to an attack of that agent, and diminishing the possibility of an attack is not achieving some kind of immunity, and not being clear with the language, is what may confuse policy makers to think we can rely on something that does not exist \citep{doi:10.1177/0890117120930536d}.

In the example we used, $R_0=2$, and, at this value, as we mentioned before, about $80\%$ of the population on average will become infected, thus, announcing that the policy to implement is based on herd immunity is similar to announce that nothing will be done. Clearly, other measures tending to reduce the contact rate will have an effect in reducing $R_0$, but the $R_0$ of COVID-19 is very high, close to 3 \citep{liu2020reproductive} and at this value, 95\% of the population will be infected. If we can reduce the contact rate in $40\%$, continuously, which is an incredible and almost impossible task, we will reduce the expected fraction of infected to 73\%, still very high. If the effort is not continuous and the contact rate is released to its natural level, then, the virus will take its toll and we will go back to the 95\% epidemic size.

\section{Bibliography}

\bibliography{/Users/carlosmh1/Dropbox/Current_Research/Literature/covid}

\end{document}